%% file: main.tex
\documentclass[preprint,12pt, a4paper]{elsarticle}

\usepackage{amssymb}
\usepackage{amsmath}
\usepackage{hyperref}

\journal{SoftwareX}

\usepackage{subcaption}
\input{listing.tex}
\graphicspath{{images/}}

\usepackage[normalem]{ulem}

\newcommand{\ArxivVersion}{0}

\begin{document}

\begin{frontmatter}



\newcommand{\jylb}[1]{\textcolor{red}{JYLB:#1}}
\newcommand{\hossein}[1]{\textcolor{blue}{Hossein:#1}}
\newcommand{\stephan}[1]{\textcolor{green}{Stéphan:#1}}

\title{Saihu: A Common Interface of Worst-Case Delay Analysis Tools for Time-Sensitive Networks}

\author{Chun-Tso Tsai\corref{cor1}}
\ead{chun-tso.tsai@epfl.ch}

\author{Seyed Mohammadhossein Tabatabaee\corref{cor2}}
\ead{hossein.tabatabaee@epfl.ch}

\author{Stéphan Plassart\corref{cor3}}
\ead{stephan.plassart@epfl.ch}

\author{Jean-Yves Le Boudec\corref{cor4}}
\ead{jean-yves.leboudec@epfl.ch}

\cortext[cor1]{Corresponding author}

\address{School of Computer and Communication Sciences, \\École Polytechnique Fédérale de Lausanne}

\begin{abstract}
   Time-sensitive networks, as in the context of IEEE-TSN and IETF-Detnet, require bounds on worst-case delays. Various network analysis tools compute such bounds; these tools are based on different methods and provide delay bounds that are all valid but may differ; furthermore, it is generally not known which tool will provide the best bound. To obtain the best possible bound, users need to implement multiple pieces of code with a different syntax for every tool, which is impractical and error-prone. To address this issue, we present Saihu, a Python interface that integrates the three most frequently used worst-case network analysis tools: xTFA, DiscoDNC, and Panco. They altogether implement six analysis methods. Saihu provides a general interface that enables defining a network in a single file and executing all tools simultaneously without any modification. Saihu further exports analysis results as formatted reports automatically and allows quick generation of certain types of networks. With its simplified steps of execution, Saihu reduces the burden on users and makes it accessible for anyone working with time-sensitive networks. An introductory video is available at \url{https://youtu.be/MiOhLay8Kr4}.
\end{abstract}


\end{frontmatter}

\input{content/introduction.tex}

\input{content/model.tex}
\input{content/tool.tex}

\input{content/software.tex}
\input{content/performance}
\input{content/impact.tex}
\input{content/conclusion.tex}

\section*{Declaration of competing interest}
The authors declare that they have no known competing financial interests or personal relationships that could have appeared to influence the work reported in this paper.





\bibliographystyle{elsarticle-num} 
\bibliography{Ref.bib}








\section*{Current code version}
\label{}


\begin{table}[!h]
\begin{tabular}{|l|p{6.5cm}|p{6.5cm}|}
\hline
\textbf{Nr.} & \textbf{Code metadata description} & \textbf{Information} \\
\hline
C1 & Current code version & v1 \\
\hline
C2 & Permanent link to code/repository used for this code version & {\small \href{https://github.com/adfeel220/Saihu-TSN-Analysis-Tool-Integration}{https://github.com/adfeel220/Saihu-TSN-Analysis-Tool-Integration}} \\
\hline
C3  & Permanent link to Reproducible Capsule & \\
\hline
C4 & Legal Code License   & MIT License \\
\hline
C5 & Code versioning system used & Git \\
\hline
C6 & Software code languages, tools, and services used & Python, Java (by DiscoDNC), lpsolve (by Panco), and CPLEX (by \textit{LUDB} option of DiscoDNC)\\
\hline
C7 & Compilation requirements, operating environments \& dependencies & Python packages: \texttt{xTFA}, \texttt{Panco}, Python packages \texttt{numpy}, \texttt{networkx}, \texttt{matplotlib}, \texttt{mdutils}, and \texttt{pulp}. Java package: \texttt{DiscoDNC}\\
\hline
C8 & If available Link to developer documentation/manual & 
\href{https://github.com/adfeel220/Saihu-TSN-Analysis-Tool-Integration/blob/main/README.md}{https://github.com/adfeel220/Saihu-TSN-Analysis-Tool-Integration/blob/main/README.md}\\
\hline
C9 & Support email for questions & \href{mailto:chun-tso.tsai@epfl.ch}{chun-tso.tsai@epfl.ch}\\
\hline
\end{tabular}
\caption{Code metadata}
\label{} 
\end{table}




\ifdefined\ArxivVersion
    \clearpage
    \appendix
    \input{content/appendix}
\fi

\end{document}

%% file: listing.tex
\usepackage{listings}

\usepackage{color}
\definecolor{gray}{rgb}{0.4,0.4,0.4}
\definecolor{darkblue}{rgb}{0.0,0.0,0.8}
\definecolor{cyan}{rgb}{0.0,0.6,0.6}
\definecolor{orange}{rgb}{0.8,0.4,0.0}
\definecolor{lightgreen}{rgb}{0.0,0.8,0.3}
\definecolor{codegreen}{rgb}{0,0.6,0}
\definecolor{codegray}{rgb}{0.5,0.5,0.5}
\definecolor{codepurple}{rgb}{0.3,0,0.52}
\definecolor{codepink}{rgb}{0.8, 0.35, 0.7}
\definecolor{lightblue}{rgb}{0.0,0.3,0.7}

\newcommand{\basicfontsize}{\footnotesize}

\lstdefinelanguage{XML}{
    moredelim=[s][\color{cyan}]{\ }{=},
    moredelim=[s][\color{black}]{>}{<},
    morestring=[b]",
    morecomment=[s]{?}{?},
    morecomment=[s]{!--}{--},
    stringstyle=\color{orange},
    identifierstyle=\color{darkblue},
    keywordstyle=\color{cyan},
    morekeywords={xmlns,version,type}
}

\newcommand{\jsonkey}{\color{darkblue}}
\newcommand{\jsonvalue}{\color{orange}}
\newcommand{\jsonnumber}{\color{cyan}}
\newcommand{\jsonbool}{\color{codegreen}}

\makeatletter
\newif\ifisvalue@json
\newif\ifisarray@jsonArray

\lstdefinelanguage{json}{
    tabsize             = 4,
    showstringspaces    = false,
    keywords            = {false,true},
    alsoletter          = 0123456789.,
    morestring          = [s]{"}{"},
    stringstyle         = \jsonkey\ifisvalue@json\jsonvalue\fi\ifisarray@jsonArray\jsonvalue\fi,
    keywordstyle        = \jsonbool,
    MoreSelectCharTable = \lst@DefSaveDef{`:}\colon@json{\enterMode@json},
    MoreSelectCharTable = \lst@DefSaveDef{`,}\comma@json{\exitMode@json{\comma@json}} \ifisarray@jsonArray\array@json{\enterMode@json}\fi,
    MoreSelectCharTable = \lst@DefSaveDef{`\{}\bracket@json{\exitMode@json{\bracket@json}},
    MoreSelectCharTable = \lst@DefSaveDef{`\{}\bracket@jsonArray{\exitMode@jsonArray{\bracket@jsonArray}},
    MoreSelectCharTable = \lst@DefSaveDef{`[}\lb@jsonArray{\enterMode@jsonArray{\lb@jsonArray}},
    MoreSelectCharTable = \lst@DefSaveDef{`]}\rb@jsonArray{\exitMode@jsonArray{\rb@jsonArray}},
    basicstyle          = \scriptsize\ttfamily
}

\newcommand\enterMode@json{%
    \colon@json%
    \ifnum\lst@mode=\lst@Pmode%
        \global\isvalue@jsontrue%
    \fi
}

\newcommand\exitMode@json[1]{#1\global\isvalue@jsonfalse}

\lst@AddToHook{Output}{%
    \ifisvalue@json%
        \ifnum\lst@mode=\lst@Pmode%
            \def\lst@thestyle{\jsonnumber}%
        \fi
    \fi
    \lsthk@DetectKeywords%
}

\newcommand\enterMode@jsonArray{%
    \lb@jsonArray%
    \ifnum\lst@mode=\lst@Pmode%
        \global\isarray@jsonArraytrue%
    \fi
}

\newcommand\exitMode@jsonArray[1]{#1\global\isarray@jsonArrayfalse}

\lst@AddToHook{Output}{%
    \ifisarray@jsonArray%
        \ifnum\lst@mode=\lst@Pmode%
            \def\lst@thestyle{\jsonnumber}%
        \fi
    \fi
    \lsthk@DetectKeywords%
}

\newcommand{\lstsetblack}{
    \exitMode@json{}
    \exitMode@jsonArray{}
}
\lstdefinestyle{pythonstyling}{
language=Python,
keywords={from,import},
basicstyle=\basicfontsize\ttfamily\color{black},
morekeywords={self},              
keywordstyle=\color{codepink},
identifierstyle=\color{lightblue},
emph={MyClass,__init__},          
emphstyle=\color{darkblue},    
stringstyle=\color{orange},
commentstyle=\color{codegreen},
frame=single,                         
showstringspaces=false
}

%% file: content/introduction.tex
\section{Introduction}
\label{sec: introduction}

Time-sensitive networks are used for real-time applications in various automation systems and require bounds on worst-case delays (cf. IEEE-TSN and IETF-Detnet). In such networks, communication flows are required to be constrained at their sources by traffic envelopes, also called arrival curves. The computation of good delay bounds is typically done using network calculus, which abstracts the service offered by a network element by a function called a service curve~\cite{le_boudec_network_2001,bouillard_deterministic_2018,nancy}. Finding the best delay bound, given the arrival curves of flows at the sources and the service curves offered by the nodes, is an NP-hard problem. 
Therefore, several methods were developed to find good delay bounds. Frequently used methods are Total Flow Analysis (TFA)~\cite{thoma2022analyse}, Separate Flow Analysis (SFA)~\cite{bouillard_deterministic_2018}, Pay Multiplexing Only Once (PMOO)~\cite{PMOO}, 
Least Upper Delay Bound (LUDB)~\cite{scheffler2021fifo}, Polynomial Linear Programming (PLP), and Exponential Linear Programming (ELP)~\cite{bouillard2022tradeoff}.
All methods provide valid delay bounds but differ in their design and implementation, and it is not trivial to identify the best, smallest bound among them. Therefore, for every application case, it is interesting to compare different methods and find the smallest delay bound.

The existing worst-case delay analysis tools, such as xTFA~\cite{thoma2022analyse}, DiscoDNC~\cite{scheffler2021fifo,bondorf2014discodnc}, and Panco~\cite{bouillard2022tradeoff} (see~\cite{listtoolWiki} for more tools), support some of the frequently used methods. These tools altogether cover most of the widely recognizable methods within the community. As of today, despite the great potential of utilizing multiple tools, users must implement multiple pieces of code with different syntaxes for each of them, which is impractical and error-prone. 

\begin{figure}[tbh]
\centering
\includegraphics[width=0.85\linewidth]{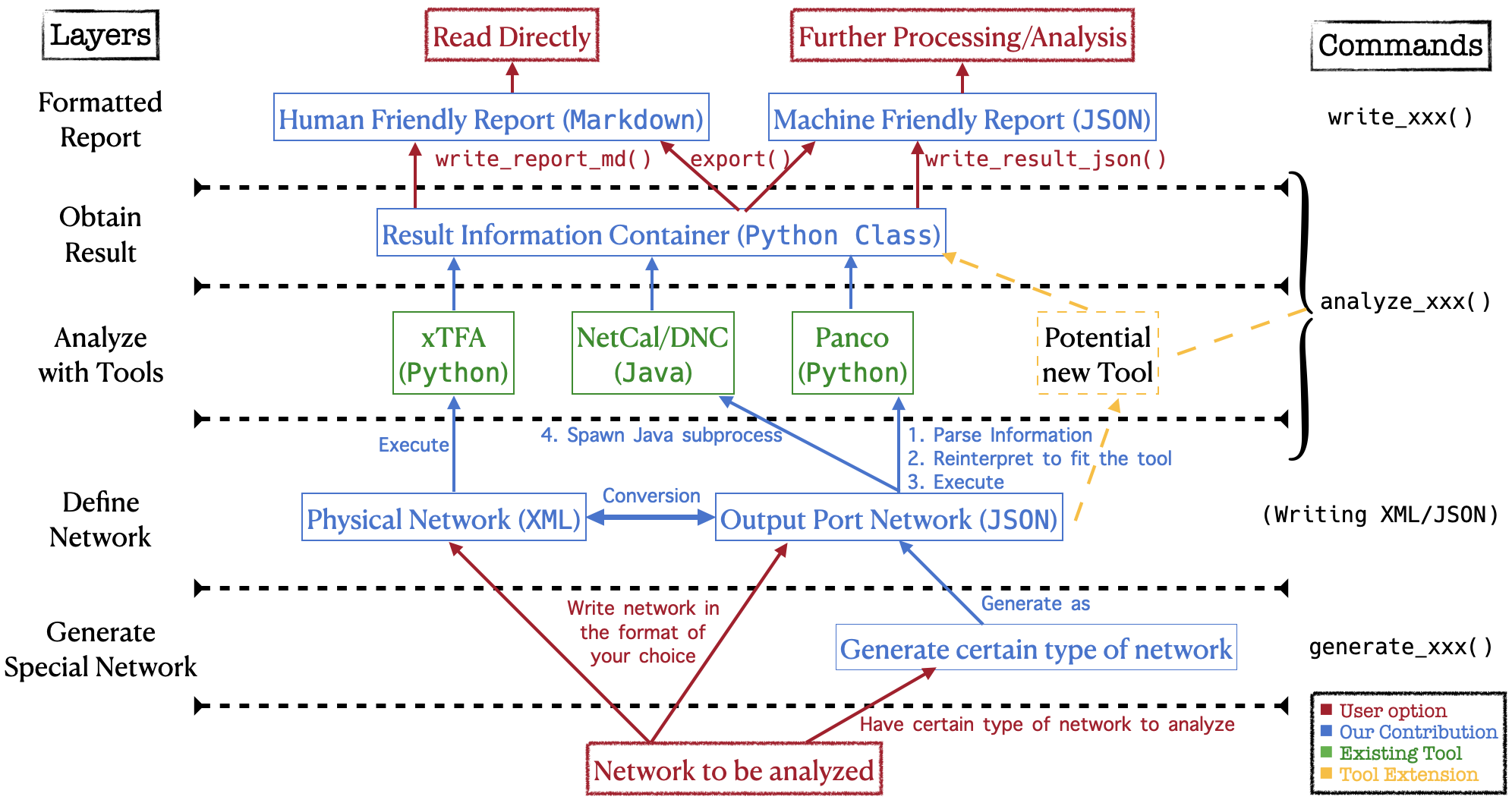}
\caption{Data flow of Saihu}
\label{fig: pipeline}
\end{figure}

We present Saihu, \textbf{S}uperimposed worst-case delay \textbf{A}nalysis \textbf{I}nterface for \textbf{H}uman-friendly \textbf{U}sage, to simplify the whole process.
Users can execute analyses and compare the results from each tool easily with a single interface and simple commands. Saihu provides a general interface that enables defining the networks in one XML or JSON file and executing all tools simultaneously without any modification; it automatically generates input for each tool respectively and executes the analyses on them. Saihu can produce analysis results in formatted reports and offer automatic network generation for certain types of networks. Therefore, with its straightforward syntax and ease of execution, Saihu simplifies the worst-case bounds comparisons in time-sensitive networks.
Its design is modular and supports the addition of new tools. 
Fig.~\ref{fig: pipeline} illustrates the design of Saihu with its data flow.

%% file: content/model.tex
\section{System Model}
\label{sec: system model}

\begin{figure}[thb]
    \centering
    \includegraphics[width=0.75\textwidth]{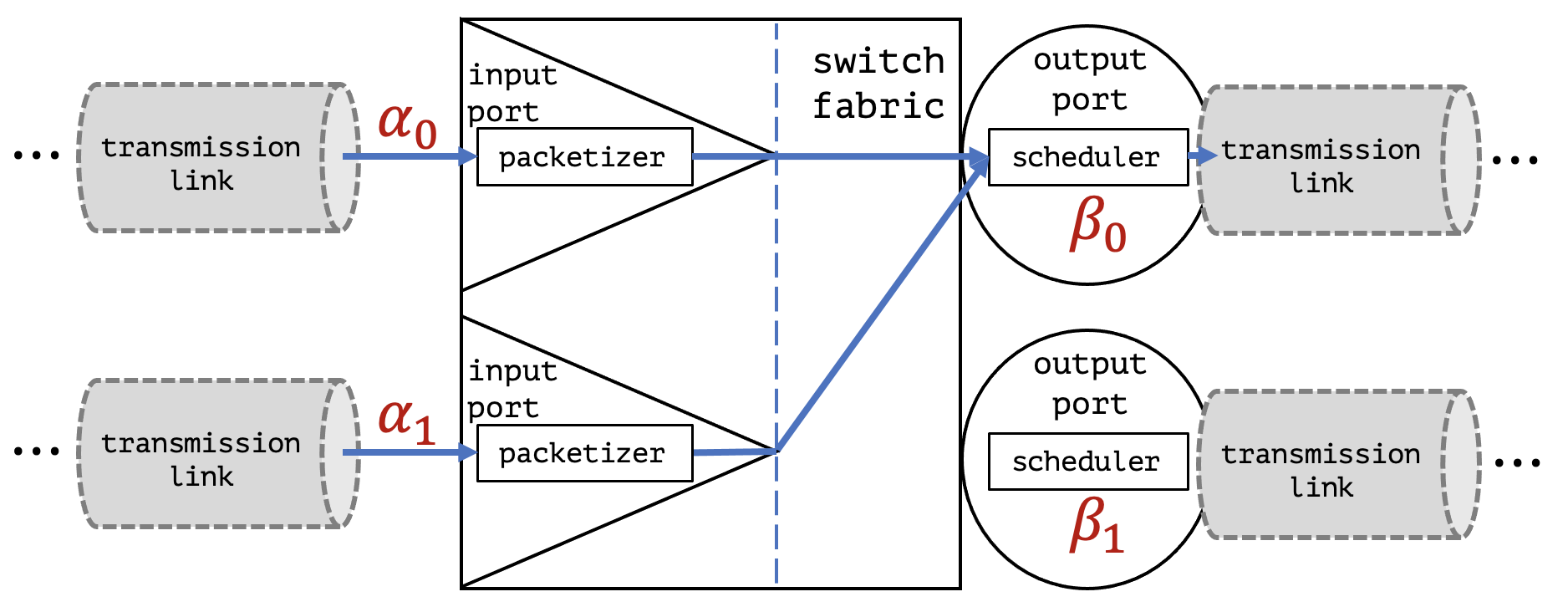}
    \caption{Device model}
    \label{fig: device}
\end{figure} 

Devices represent switches or routers that compose the network of interest; they consist of input ports, output ports, and switching fabric. Fig.~\ref{fig: device} shows one such device. Each packet enters a device via an input port and is stored in a packetizer. A packetizer releases a packet only when the entire packet is received. Then, the packet goes through a switching fabric, which transmits the packet to a specific output port based on the static route of the packet; the packet is either buffered in a FIFO (First-In-First-Out) queue and then is serialized on the output line at the transmission rate of the line or exits the network via a terminal port (i.e., a sink). 

A flow is a stream of packets generated from the same source, following the same path, and destined for the same sink.  We assume that flows are statically assigned to a path. A path consists of a source, a sequence of devices (with the corresponding input port and output port), and a sink (see Fig.~\ref{fig: physical network}). For each flow~$i$, we let $l^{\scriptsize\textrm{max}}_i$ and $l^{\scriptsize\textrm{min}}_i$ denote the maximum and minimum packet size, respectively. Every flow is constrained at the source by an arrival curve which we assume to be piece-wise linear and concave. Such an arrival curve, say $\alpha$, can be described by a collection of $m$~rates~$r_1, r_2, ..., r_m$ and bursts $b_1, b_2, ...,b_m$ such that $\alpha(t) = \min_{k=1:m}(r_kt + b_k)$; each function $t \mapsto r_kt + b_k$ is called a token-bucket function with rate~$r_k$ and burst~$b_k$. The long-term arrival rate of a flow is $\min_k r_k$. Note that the parameters $(m, r_{1:m}, b_{1:m})$ may differ for every flow. 

A flow can be either \textit{unicast} (one source, one destination) or \textit{multicast} (one source, multiple destinations). In the latter case, traffic can be split at one or several intermediate devices. 
For the tools that do not support multicast flows, we replace every multicast flow with $p$ sub-paths by $p$ unicast flows with the same arrival curve constraint at the source; this increases the traffic inside the network, and thus delay bounds that we obtain are valid but might be less good than those obtained by tools that natively support multicast flows.  

The service offered to the aggregation of all flows of interest at an output port is represented by a service curve, which we assume to be piece-wise linear and convex. Such a service curve, say $\beta$, can be described by a collection of $n$~rates $R_1, R_2, ..., R_n$ and latencies $T_1, T_2, ...,T_n$ such that $\beta(t) = \max_{k=1:n}(R_k[t - T_k]^+)$, with the notation $[x]^+ = \max(x,0)$; each function $t \to R_k[t - T_k]^+$ is called a rate-latency function with rate~$R_k$ and latency~$T_k$. The long-term service rate of the output port is defined as $\max_k R_k$. Note that the parameters $(n, R_{1:n}, T_{1:n})$ may differ at every output port.

We assume that the network is locally stable, namely, at every output port, the aggregate long-term arrival rate (equal to the sum of the long-term arrival rates of all flows using the output port) is less than the long-term service rate. This is a necessary condition for the existence of finite delay bounds; it is also sufficient in feed-forward networks, but not in networks that have cyclic dependencies \cite[Chapter 12]{bouillard_deterministic_2018}.

%% file: content/tool.tex
\section{Included Tools}
\label{sec: included tools}

Saihu currently includes 3 tools:  xTFA, DNC, and Panco. 
Fig.~\ref{fig: supported methods} summarizes supported methods for each tool.

\begin{center}
    \begin{tabular}{|c|c|c|c|}
        \hline
        Method\textbackslash Tool & DNC & xTFA & Panco \\
        \hline\hline
        TFA  & V & V & V \\
        SFA  & V &   & V \\
        PLP  &   &   & V \\
        ELP  &   &   & V \\
        PMOO & V &   &   \\
        LUDB & V &   &   \\
        \hline
    \end{tabular}
    \captionof{table}{Supported methods are marked with a ``V''
    }
    \label{fig: supported methods}
\end{center}

\begin{itemize}
    \item \textbf{xTFA}~\cite{thoma2022analyse} is developed in Python and supports a more advanced TFA. For its input, an XML file describes the network (cf. Sec.~\ref{sec: physical network xml}). 
    xTFA supports analyzing networks with cyclic dependency and multicast flows.

    \item \textbf{DiscoDNC}~\cite{bondorf2014discodnc} is developed in Java and partially uses linear programming with \textit{CPLEX}~\cite{cplex2009v12} for LUDB. It supports TFA, SFA, PMOO, and LUDB. A network is defined through its own Java classes. Saihu uses the information from an output port network to create a network in DNC syntax internally. Moreover, with DNC, one cannot manually set shaping with FIFO multiplexing but only with arbitrary multiplexing. Also, DNC does not support networks with cyclic dependency and does not support multicast flows (see Section~\ref{sec: system model}).

    \item \textbf{Panco}~\cite{bouillard2022tradeoff} is developed in Python and uses linear programming. So, it requires \textit{lpsolve}~\cite{lpsolve} to execute TFA, SFA, PLP, and ELP. A network is described with its own Python classes. Saihu internally creates the network in Panco syntax from the information of an output port network. All methods of Panco except for ELP support networks with cyclic dependencies. Panco  does not support multicast flows (see Section~\ref{sec: system model}).
\end{itemize}

%% file: content/software.tex
\section{Software Description}
\label{sec: software description}

Saihu's analyses are done in 3 steps: describe a network to be analyzed (Sec.~\ref{sec: network description file}); execute analyses with selected tools (Sec.~\ref{sec: tool usage}); and export analysis reports back to the user (Sec.~\ref{sec: analysis reports}). 

\subsection{Network Description File}
\label{sec: network description file}

\begin{figure}
\centering
\begin{subfigure}[b]{0.63\textwidth}
    \centering
    \includegraphics[width=\linewidth]{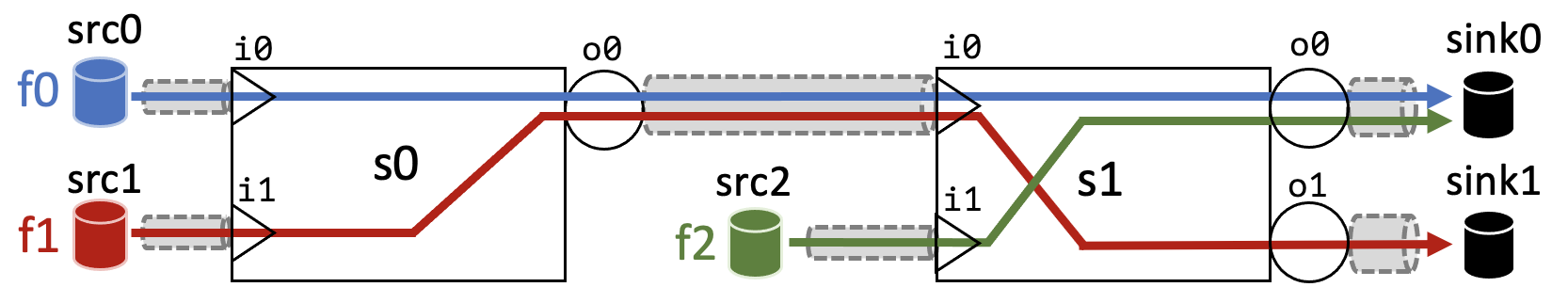}
    \caption{
    Physical Network. 
    }
    \label{fig: physical network}
\end{subfigure}
\hfill
\begin{subfigure}[b]{0.35\textwidth}
    \centering
    \includegraphics[width=0.7\linewidth]{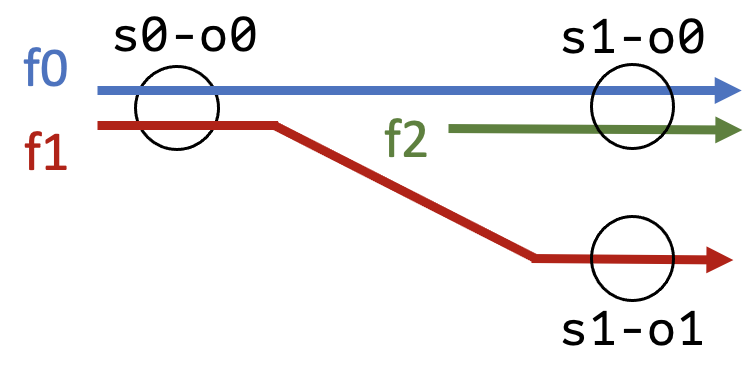}
    \caption{Output Port Network.}
    \label{fig: output port network}
\end{subfigure}
\caption{Physical and output port network examples}
\label{fig: physical and op network}
\end{figure}

Saihu allows the user to write a network in either a \textit{physical network} or an \textit{output port network} format. Examples are shown in Fig.~\ref{fig: physical and op network}. Briefly speaking, a physical network represents the physical connections between multiple switches and stations and flows that travel through different input and output ports of switches; it represents the view of a real-world network. On the other hand, as a physical network includes more than enough information, we provide the output port network format as a simplified form to define a network. As the queuing is only on output ports even if we provide full network information, we only describe output ports as service units instead of the entire device. 

To allow easy access to the tool, we provide both physical and output port networks as available input file forms. People may prefer to directly write in the physical network format to avoid the translation to an output port format, and some people may prefer the output port network to write a network concisely.
While xTFA takes a physical network as an XML file and the others parse from an output port network as a JSON file, one can choose the format they prefer to define a network as Saihu automatically converts a file when needed. 

\subsubsection{Option 1: Physical Network in XML}
\label{sec: physical network xml}
A physical network is written as an XML file in the same format as in xTFA, and at least contains \textbf{General network information}, \textbf{Servers}, \textbf{Links}, and \textbf{Flows}. 
Listing.~\ref{lst: xml example} demonstrates a minimal example.

\begin{lstlisting}[language=XML,caption={Example of a physical network representation},
label={lst: xml example}]
<network name="demo" technology="FIFO+IS" minimum-packet-size="50B"/>
<station name="src0"/>
<switch name="s0" service-latency="10us" service-rate="4Mbps"/>
<station name="sink0"/>
<link name="src0-s0" from="src0" to="s0" fromPort="o0" toPort="i0"/>
<link name="s0-sink0" from="s0" to="sink0" fromPort="o0" toPort="i0" transmission-capacity="10Mbps"/>
<flow name="f0" arrival-curve="leaky-bucket" lb-burst="10B" lb-rate="10kbps" maximum-packet-size="50B" source="src0">
    <target>
        <path node="s0"/>
        <path node="sink0"/>
    </target>
</flow>
\end{lstlisting}

First, one \texttt{network} element defines the general network information as its attributes: the network's name (\texttt{name}), analysis parameters concatenated by the plus sign (\texttt{technology}, \texttt{IS} stands for \textit{Input Shaping}), and optionally some default value (e.g. \texttt{minimum-packet-size}) across the network. 

Second, a server can be either a \texttt{station} or a \texttt{switch} representing a physical node. Although they are physically different, they both serve as service-providing devices in our tools as mentioned in Sec.~\ref{sec: system model}, or sources/sinks of a data flow. The service parameters \texttt{service-latency} and \texttt{service-rate} define a default service curve for all the output ports on this device.

Third, a \texttt{link} connects two devices. Saihu tools consider output ports as processing units, so the physical link must be defined \texttt{from} a physical node \texttt{to} another node with the input and output ports used by the link. Namely, it goes from an output port of one server to an input port of another server.
Since a link directly attaches to an output port, users can define service via a link. The \texttt{transmission-capacity} of the link can also be specified to consider line shaping. Without defined values, the system will apply the default values defined at the upper levels (\texttt{switch/station} or \texttt{network}).

Finally, a \texttt{flow} element defines a flow. Flow paths are surrounded by \texttt{target} elements, where each node it traverses is listed as a \texttt{path} element with its \texttt{node} attribute indicating the name of the physical node. In this format, a multicast flow is possible and is obtained by defining multiple \texttt{target} elements within the same flow.
A token-bucket curve at the \texttt{source} is defined by \texttt{arrival-curve}, \texttt{lb-burst}, and \texttt{lb-rate} keywords. Packetization is considered with maximum and minimum packet sizes. Saihu analyzes all the output ports in the order of the flow path.

\subsubsection{Option 2: Output Port Network in JSON}
\label{sssec: json}

Output port format is designed by the authors to write the network concisely. The file contains at least \textbf{General network information}, \textbf{Servers}, and \textbf{Flows}. An example is shown as Listing~\ref{lst: json example}.

First, a \texttt{network} object defines the general network information, the default values, and units throughout the network.

Second, \texttt{servers} defines all servers as an array. The parameters can be either a \textit{string} as a number followed by a unit, e.g. \texttt{"10us"} for 10 microseconds; or a \textit{number} that uses the predefined unit.
The service curve is taken as the maximum of all rate-latency curves defined in \texttt{service\_curve} (see Sec.~\ref{sec: system model}). 
Each rate-latency curve is described by a pair of rate and latency values with the same index. For example, the service curve of server \texttt{s0-o0} is derived from 2 rate-latency curves: the first has a 10 microseconds latency and 4 megabits per second rate, and the second has a 1000 microseconds latency and 50 megabits per second rate.

Notice that in the output port network format, we use \textit{graph-induced-by-flows} as the network topology instead of manually defining links. A link between two servers exists only when at least one flow crosses these two servers consecutively. Therefore, the link's transmission capacity attached to an output port is directly defined on a server with the keyword \texttt{capacity}.

\begin{lstlisting}[float,language=json,caption={Network information with default values},label={lst: json example}]
{
    "network": {
        "name": "demo",
        "multiplexing": "FIFO",
        "rate_unit": "Mbps"
    },
    "servers": [
        {
            "name": "s0-o0",
            "service_curve": {
                "latencies": ["10us", "1ms"],
                "rates": [4, "50Mbps"]
            },
            "capacity": "200Mbps"
        }
    ],
    "flows": [
        {
            "name": "f0",
            "path": ["s0-o0"],
            "arrival_curve": {
                "bursts": ["10B", "2kB"],
                "rates": ["10kbps", 0.5]
            },
            "max_packet_length": "50B",
        }
    ]
}
\end{lstlisting}
\lstsetblack

Finally, \texttt{flows} represents the flows as an array. Each flow is defined by a \texttt{path} as an array of servers, and an \texttt{arrival\_curve} at its source.
The arrival curve is defined as the minimum of the multiple token-bucket curves with each pair of burst and rate values represents a token-bucket curve (see Sec.~\ref{sec: system model}). For example, \texttt{f0}'s arrival curve is composed of a token-bucket curve of burst 10 bytes and rate 10 kilobits per second, and a curve of burst 2 kilobytes and rate 0.5 megabits per second.

All flows written in the output port network format are assumed to be unicast flows. When being converted from a physical network with multicast flows, it separates the paths into multiple unicast flows with the same source and arrival curve (see Section \ref{sec: system model}).

\subsection{Tool Usage}
\label{sec: tool usage}

Saihu analysis execution can be done via the command line tool \texttt{main.py} or by importing the interface from \texttt{interface.py}. We demonstrate the simplest way to analyze a network file, say \textit{demo.json}, with both possibilities. Listing~\ref{lst: cmd example} and Listing~\ref{lst: package example} show two ways to analyze \textit{demo.json} with all the tools and methods available inside Saihu.

\begin{lstlisting}[float,language=bash,caption={Use Saihu as command line tool},label={lst: cmd example}]
python main.py demo.json -a
\end{lstlisting}

\begin{lstlisting}[float,style=pythonstyling,caption={Use Saihu as package},label={lst: package example}]
from saihu.interface import TSN_Analyzer
analyzer = TSN_Analyzer("demo.json")
analyzer.analyze_all()
analyzer.export("demo")
\end{lstlisting}

Manual selection of tools or methods is possible via specifying different flags in the command line interface, or using different function names as \texttt{analyze\_xxx}, with the \texttt{analyze\_all} in the above example representing analyzing with all tools.

\subsection{Analysis Reports}
\label{sec: analysis reports}

\begin{figure}[!tbh]
\centering
\begin{subfigure}[t]{0.72\textwidth}
    \centering
    \includegraphics[width=\linewidth]{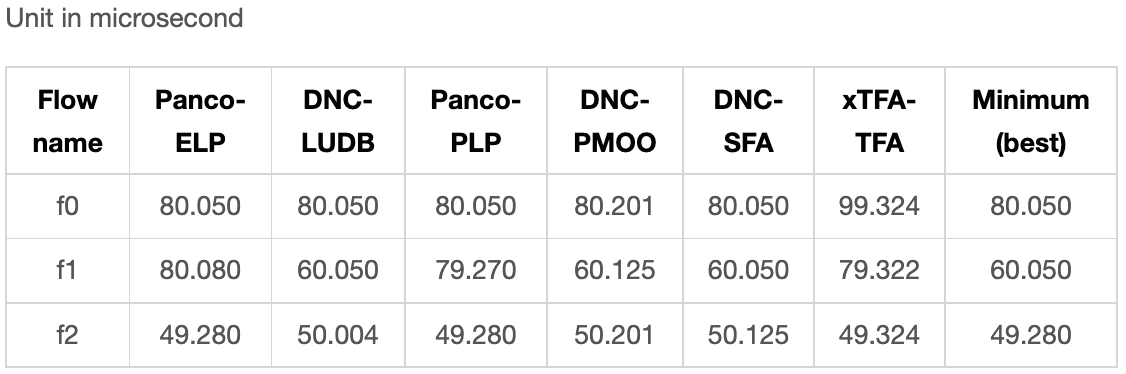}
    \caption{Flow end-to-end delay.}
    \label{fig: e2e delay}
\end{subfigure}
\hfill
\begin{subfigure}[t]{0.26\textwidth}
    \centering
    \includegraphics[width=\linewidth]{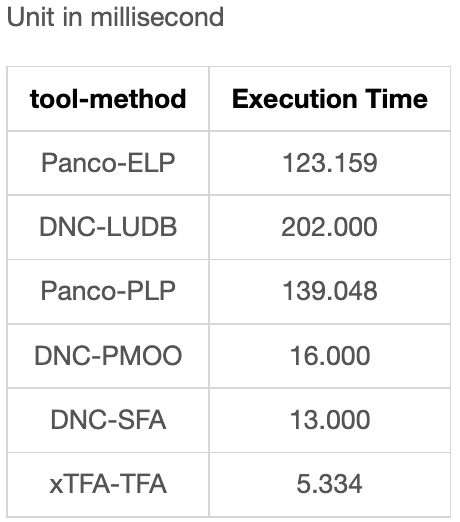}
    \caption{Execution time.}
    \label{fig: exec time}
\end{subfigure}
\caption{Human-friendly Markdown report
}
\label{fig: human friendly report}
\end{figure}

Saihu can generate 2 kinds of reports:
\begin{enumerate}
    \item A \textit{human-friendly report} is generated as a Markdown file that gives the per-flow end-to-end delay, per-server delay, and execution time for each tool. The delay bounds are presented in tables where each row is a flow or a server, and each column is a method executed by a tool. The last column contains the minimum result obtained in the current round of analysis. Fig.~\ref{fig: human friendly report} demonstrates an example of per-flow end-to-end delay and execution time as a reference.

    The report also contains some reminders about the user inputs: network topology using the graph-induced-by-flows (Sec.~\ref{sssec: json}), flow paths, and link utilization by nodes. Link utilization is defined as the ratio between the aggregated arrival rate at a node and its service rate.

    \item A \textit{machine-friendly report} is written in JSON format for easy parsing from other programs. It stores the execution outputs, i.e. the per-flow end-to-end delay, per-server delay, and execution time. An example is shown in Listing~\ref{lst: machine friendly report}. 
    Note that the numbers in a human-friendly report are always rounded to 3 decimal digits while there's no such rounding for a machine-friendly report. As a result, one should read the machine-friendly report if they require a very precise result.

\end{enumerate}

\begin{lstlisting}[language=json,caption={JSON report},label={lst: machine friendly report}]
{
    "name": "demo",
    "flow_e2e_delay": {
        "f0": {
            "xTFA_TFA": 99.32394489448944,
            "Panco_PLP": 80.05,
            "Panco_ELP": 80.05,
            "DNC_SFA": 80.0501253132832,
            "DNC_PMOO": 80.20050125313283,
            "DNC_LUDB": 80.0501253132832
        },
...
    "execution_time": {
        "xTFA_TFA": 5.716085433959961,
        "Panco_PLP": 147.26519584655762,
        "Panco_ELP": 129.76408004760742,
        "DNC_SFA": 12.0,
        "DNC_PMOO": 9.0,
        "DNC_LUDB": 172.0
    },
    "units": {
        "flow_delay": "us",
        "server_delay": "us",
        "execution_time": "ms"
    }
}
\end{lstlisting}
\lstsetblack

%% file: content/performance.tex
\section{Case Studies}
\label{sec:caseStudies}

\ifdefined\ArxivVersion
We apply Saihu to common network topologies: Families of tandem, ring, and mesh networks, used for numerical evaluation in the Panco research~\cite{bouillard2022tradeoff}. Also, it is applied to an industrial-size network that is a test configuration provided by Airbus \cite{Grieu2004AnalyseE} and is used as a benchmark in several research papers \cite{tabatabaee2021deficit, plpdrr, charara2006methodsAFDX}. Please see~\ref{sec: numerical results}.

This helped us develop Saihu and ensure its applicability to a variety of cases. Note that the obtained results are not the main purpose of this paper, but are only provided for an interested reader in~\ref{sec: numerical results}. Specifically, the comparison of delay bounds obtained by different methods is not the main purpose of this paper. Rather, Saihu significantly reduces the effort to make such a comparison by providing a general interface to apply three frequently used worst-case delay analyses in a single shot. Hence, it solves the non-trivial problem of finding the best, smallest delay among the supported methods.

\fi

\ifdefined\SoftwareXVersion
{\color{blue}
We apply Saihu to common network typologies: Families of tandem, ring, and mesh networks, used for numerical evaluation in the Panco research~\cite{bouillard2022tradeoff}. Also, it is applied to an industrial-size network that is a test configuration provided by Airbus \cite{Grieu2004AnalyseE} and is used as a benchmark in several research papers \cite{tabatabaee2021deficit, plpdrr, charara2006methodsAFDX}. Please see Appendix A of the technical report~\cite{tsai2023saihucommoninterfaceworstcase}.

This helped us to develop Saihu and to ensure its applicability to a variety of cases. Note that the obtained results are not the main purpose of this paper, but are only provided for an interested reader in Appendix A of the technical report~\cite{tsai2023saihucommoninterfaceworstcase}. Specifically, the comparison of delay bounds obtained by different methods is not the main purpose of this paper. Rather, Saihu significantly reduces the effort to make such a comparison by providing a general interface to apply three frequently used worst-case delay analyses in a single shot. Hence, it solves the non-trivial problem of finding the best, smallest delay among the supported methods.

}
\fi

%% file: content/impact.tex
\section{Impact}
\label{sec: impact}

Worst-case delay analysis tools are essential to make theoretical network studies applicable to real-world problems. With each tool specializing in particular methods, the aggregation of these tools has great potential to find the best delay bound of a network. Saihu makes executing and comparing these analysis tools easy by encapsulating them into a single interface while hiding the complexity within. It turns a previously tedious multi-platform programming task into a straightforward ``write your problem down and press run." By simplifying the user's action to analyze a network, Saihu not only eases the work of existing researchers but also makes these tools accessible to a broader audience by requiring a more approachable level of programming skill. 

Saihu also serves as a platform for potential network analysis tool developers due to its cross-platform design. Since Saihu provides unified input/output formats, one can extend Saihu to include more tools by parsing one of the unified inputs and feeding the analysis results into the unified output class. Future developers can bridge their new tool to Saihu and allow existing Saihu users to access their tool easily without forcing people to learn a new framework. 

%% file: content/conclusion.tex
\section{Conclusion and Extension}

We presented Saihu, a Python interface capable of executing multiple network analysis tools easily. It allows users to define a network and retrieve the analysis results for multiple tools. Its unified input and output layers enable the automation of all the tedious re-interpretation of a network for each tool. Such design not only simplifies many mechanical procedures previously hindering network researchers' works but also helps publicize these useful tools to more potential users due to its simplicity.

%% file: content/appendix.tex
\section{Numerical Results}
\label{sec: numerical results}

We apply Saihu to common network typologies. We provide the numerical results we obtained for these networks. This shows the applicability of Saihu to a variety of cases. Note that the obtained results are not the main purpose of this paper, but are provided for an interested reader.

\subsection{Interleave Tandem Network}

\begin{figure}[htb]
    \centering
    \includegraphics[width=0.7\linewidth]{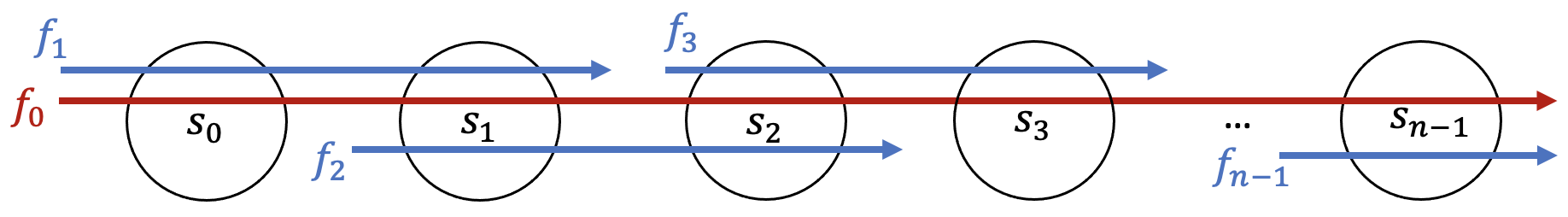}
    \caption{Interleave tandem network}
    \label{fig: interleave}
\end{figure}

An interleave tandem network is illustrated in Figure~\ref{fig: interleave}. Consider a network of $n$ servers, indexed from~$0$ to~$n-1$.
An interleave tandem network has all its servers chained in a line. One flow~$f_0$ goes through all servers from~$s_0$ to~$s_{n-1}$.  The flow~$f_i$ is $s_{i-1} \rightarrow s_{i}$ for $i \in [1,n-1]$. Every flow has the same arrival curve and the same maximum packet length at the source, defined by function arguments \texttt{burst}, \texttt{arrival\_rate}, and \texttt{max\_packet\_length}. Every server has the same service curve and transmission capacity, defined by \texttt{latency}, \texttt{service\_rate}, and \texttt{capacity}.

\begin{figure*}[thb]
\centering
\begin{subfigure}[b]{0.75\textwidth}
    \centering
    \includegraphics[width=\linewidth]{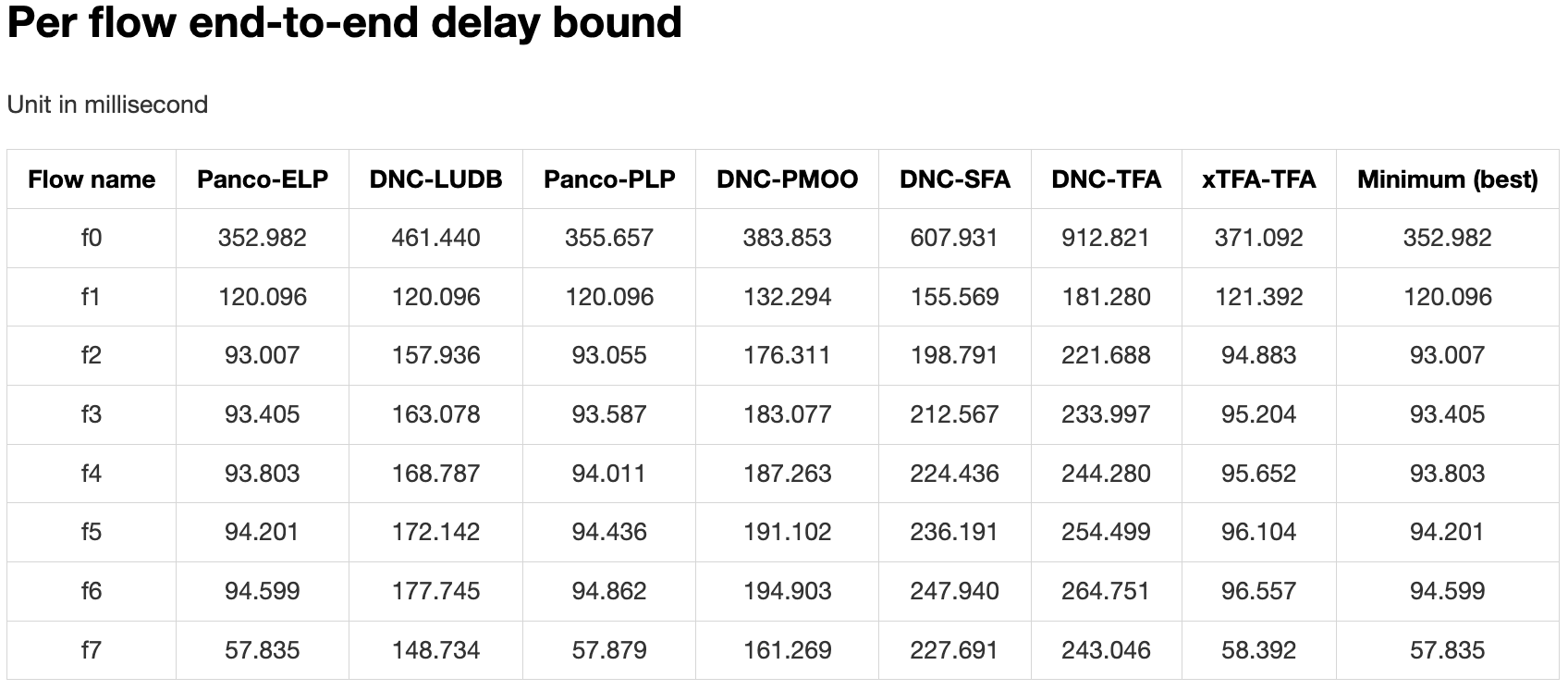}
    \caption{Flow end-to-end delay}
    \label{fig: interleave8 e2e delay}
\end{subfigure}
\hfill
\begin{subfigure}[b]{0.2\textwidth}
    \centering
    \includegraphics[width=\linewidth]{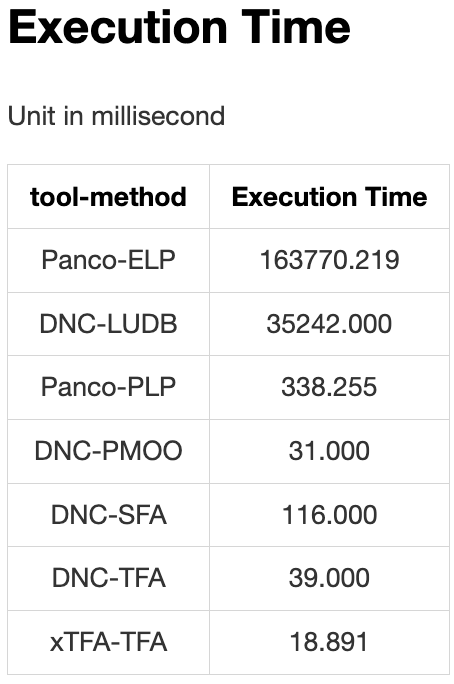}
    \caption{Execution time}
    \label{fig: interleave8 time}
\end{subfigure}
\caption{Analysis report of interleave tandem network with 8 servers}
\label{fig: interleave8 report}
\end{figure*}

We apply Saihu to an interleave tandem network of 8 servers with the parameters shown in Table~\ref{tab: interleave experiment parameters}; see Figure~\ref{fig: interleave8 report}. Due to the limitations of each method, the interleave tandem network is the only type of network in our examples that is compatible with all the methods. Such restriction does not relate to Saihu. Rather, it is the complexity of ELP. One can already see it in Figure~\ref{fig: interleave8 time}. ELP requires minutes to run while most methods return their analysis results within several milliseconds. The choice of applying ELP to networks with only less than $10$~nodes is also suggested in the original paper~\cite{bouillard2022tradeoff}.

\begin{table}[h]
    \centering
    \begin{tabular}{c|c|c|c|c|c}
        burst & arrival rate & max packet size & latency & service rate & capacity \\
        \hline
        32 bits & 64 bps & 256 bits & 10 ms & 1 kbps & 1 kbps
    \end{tabular}
    \caption{Parameters of the example interleave tandem network with 8 servers}
    \label{tab: interleave experiment parameters}
\end{table}

Even with such a small-scale network, the delay-bound estimations and the runtimes of individual methods can noticeably differ. Such an example shows the importance of Saihu, where the researchers can try out all the available methods to find the best bound relatively easily.

\subsection{Ring Network}

\begin{figure}[htb]
    \centering
    \includegraphics[width=0.3\linewidth]{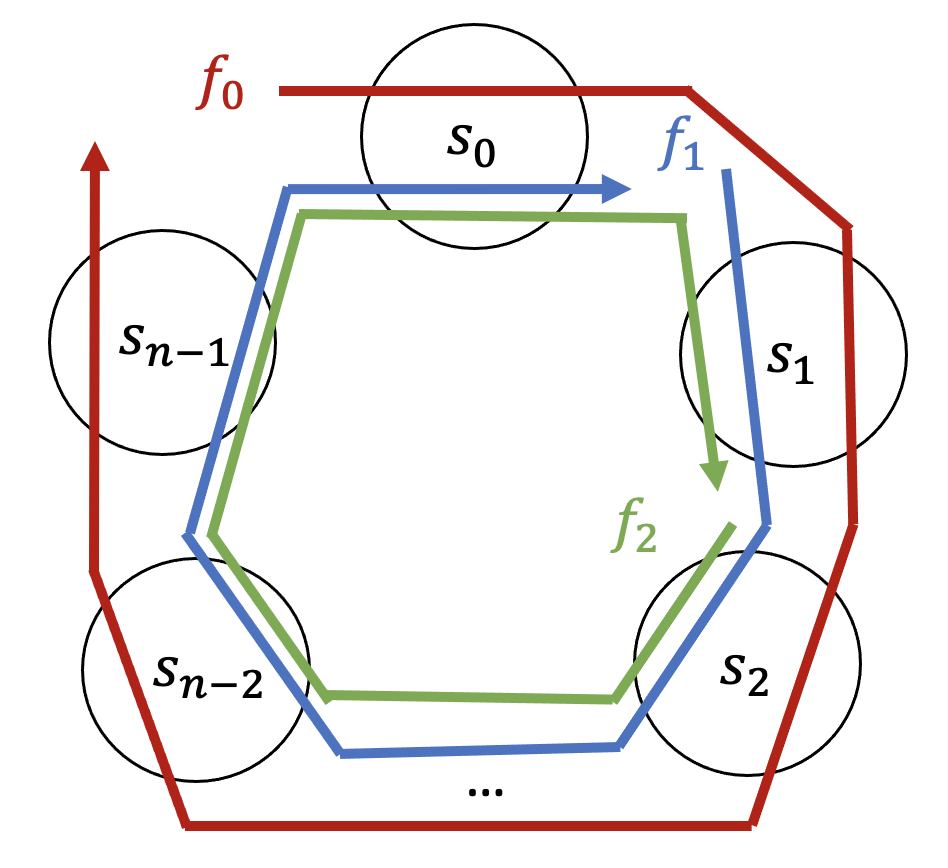}
    \caption{Ring network}
    \label{fig: ring}
\end{figure}

A ring network is illustrated in Figure~\ref{fig: ring}. There are $n$~flows and $n$~servers. The path of flow~$i$ is $s_i \rightarrow s_{i+1} \rightarrow \cdots \rightarrow s_{i+n-1\mod n}$ for $0 \leq i \leq n-1$. A ring network is completely symmetrical with identical flows and servers. An arrival curve of every flow is defined by \texttt{burst}, \texttt{arrival\_rate}, and \texttt{max\_packet\_length}. Also, a service curve of every server is defined by \texttt{latency}, \texttt{service\_rate}, and \texttt{capacity}.

\begin{figure*}[thb]
\centering
\begin{subfigure}[b]{0.47\textwidth}
    \centering
    \includegraphics[width=\linewidth]{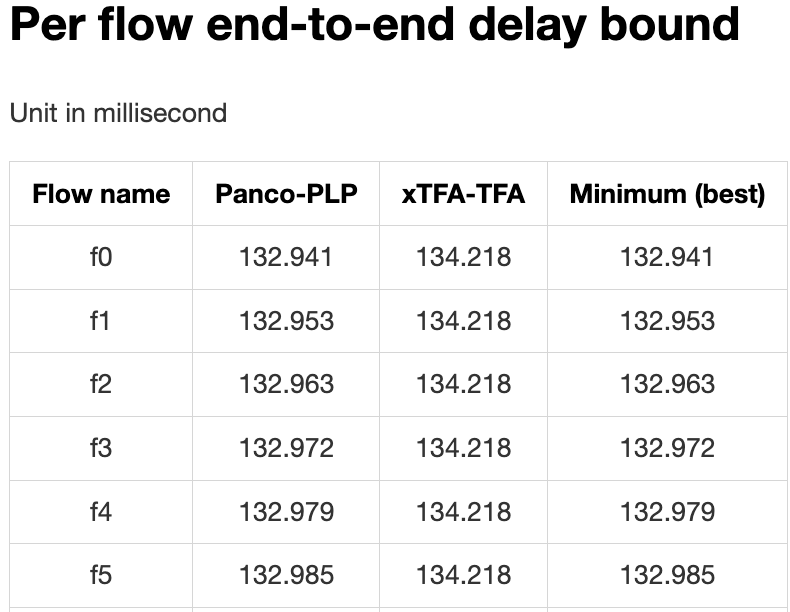}
    \caption{Flow end-to-end delay (only the first 5 flows)}
    \label{fig: ring e2e delay}
\end{subfigure}
\begin{subfigure}[b]{0.3\textwidth}
    \centering
    \includegraphics[width=\linewidth]{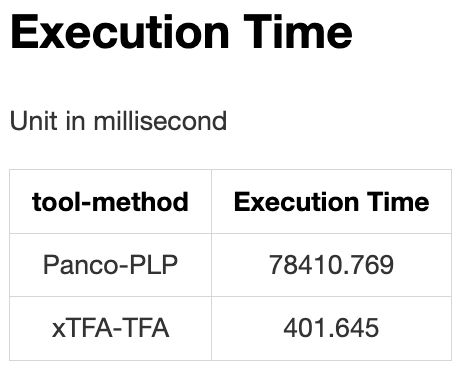}
    \caption{Execution time}
    \label{fig: ring time}
\end{subfigure}
\caption{Analysis report of a ring network with $10$ servers}
\label{fig: ring report}
\end{figure*}

We apply Saihu to a ring network of $10$~servers with the parameters shown in Table~\ref{tab: ring experiment parameters}; see Figure~\ref{fig: ring report}. As the ring network has cyclic dependencies, we can only apply the methods PLP and xTFA.

\begin{table}[h]
    \centering
    \begin{tabular}{c|c|c|c|c|c}
        burst & arrival rate & max packet size & latency & service rate & capacity \\
        \hline
        32 bits & 64 bps & 256 bits & 10 ms & 10 kbps & 10 kbps
    \end{tabular}
    \caption{Parameters of the example ring network with $10$ servers}
    \label{tab: ring experiment parameters}
\end{table}

\subsection{Mesh Network}

\begin{figure}[htb]
    \centering
    \includegraphics[width=0.4\linewidth]{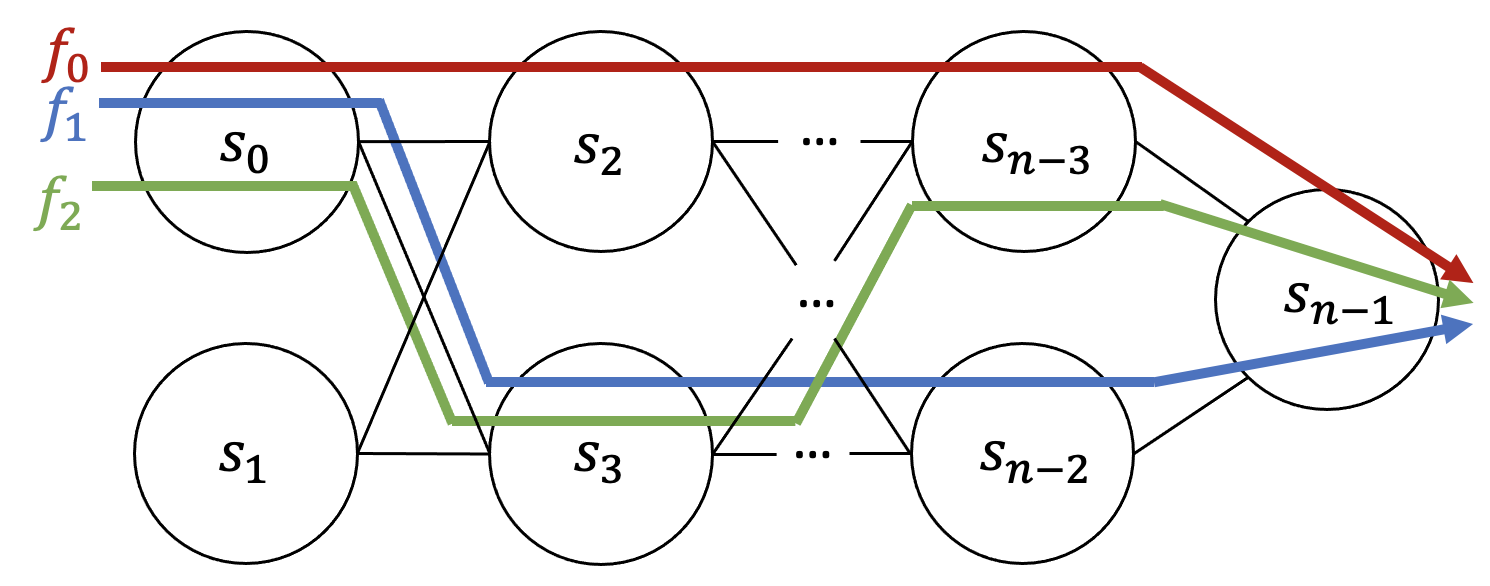}
    \caption{Mesh network. Only parts of the flows from~$s_0$ are shown. There is a flow for every possible path from~$s_0$ or~$s_1$ to~$s_{n-1}$.}
    \label{fig: mesh}
\end{figure}

A mesh network is illustrated in Figure~\ref{fig: mesh}. All flows start from either~$s_0$ or~$s_1$. The flows go through all $2^{(n-1)/2}$ possible combinations of servers towards the right, e.g. $s_0 \rightarrow s_2 \rightarrow \cdots$ and $s_1 \rightarrow s_2 \rightarrow \cdots$ are both in the network. All servers have the same service curve and capacity except~$s_{n-1}$, that has the doubled service rate. All flows have identical service curves and maximum packet length.

\begin{figure*}[thb]
\centering
\begin{subfigure}[b]{0.75\textwidth}
    \centering
    \includegraphics[width=\linewidth]{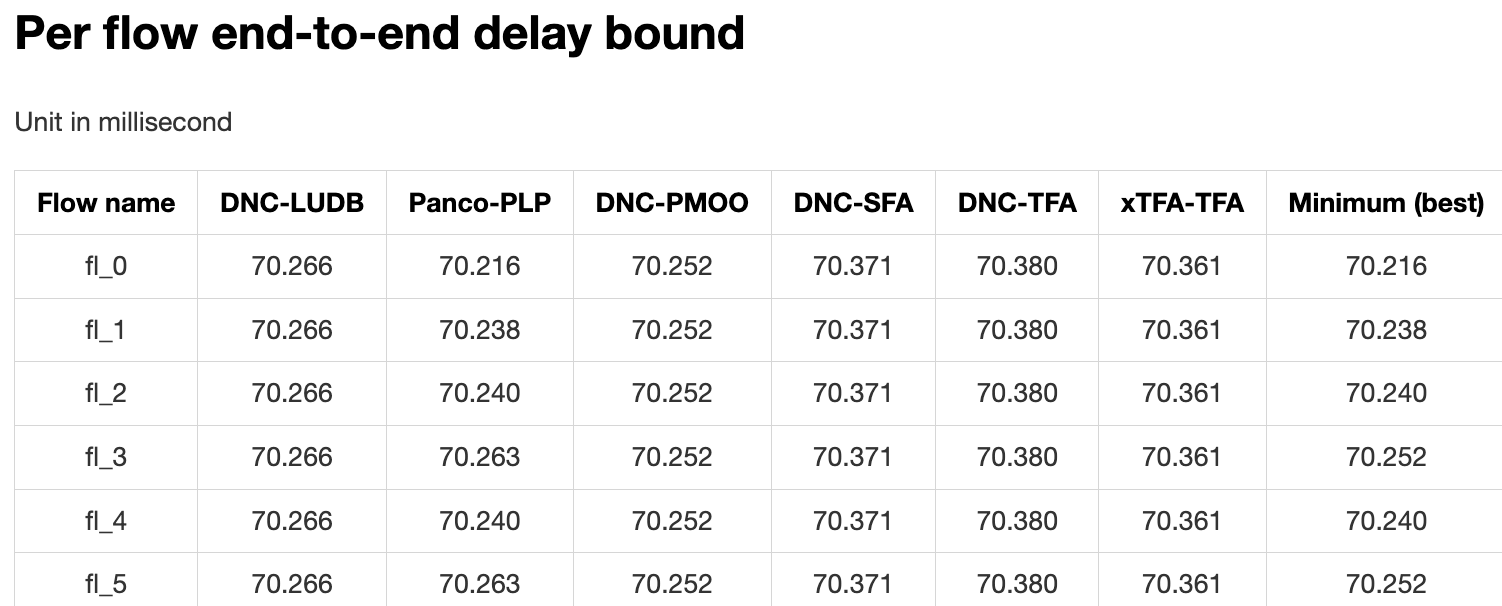}
    \caption{Flow end-to-end delay (only the first 5 flows)}
    \label{fig: mesh e2e delay}
\end{subfigure}
\begin{subfigure}[b]{0.22\textwidth}
    \centering
    \includegraphics[width=\linewidth]{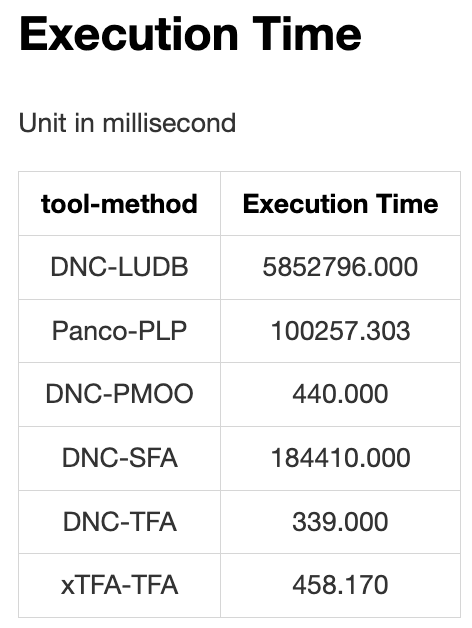}
    \caption{Execution time}
    \label{fig: mesh time}
\end{subfigure}
\caption{Analysis report of a mesh network with $13$ servers}
\label{fig: mesh report}
\end{figure*}

We apply Saihu to a mesh network of $13$~servers with the parameters shown in Table~\ref{tab: mesh experiment parameters}. We do not apply ELP because the number of flows is too large to handle. One can see the report in Figure~\ref{fig: mesh report}. Note that the complexity of SFA, LUDB, and PLP can still become intractable in a small-size network.

\begin{table}[h]
    \centering
    \begin{tabular}{c|c|c|c|c|c}
        burst & arrival rate & max packet size & latency & service rate & capacity \\
        \hline
        32 bits & 64 bps & 256 bits & 10 ms & 20 Mbps & 200 Mbps
    \end{tabular}
    \caption{Parameters of the example mesh network with $13$ servers}
    \label{tab: mesh experiment parameters}
\end{table}

\subsection{Industrial-Sized Network}
\label{sec: fixed topology random}

\begin{figure}[htb]
    \centering
    \includegraphics[width=0.5\linewidth]{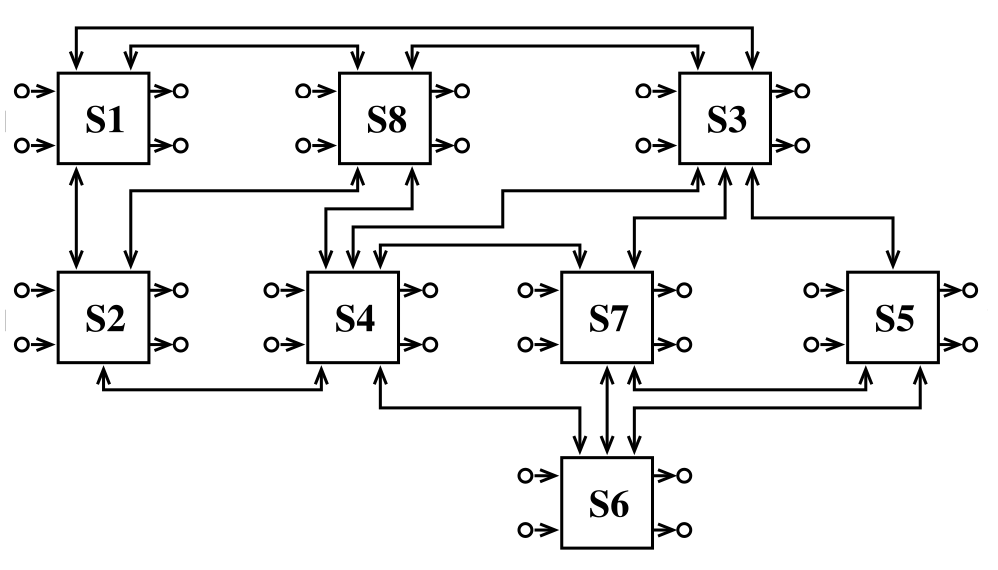}
    \caption{Industrial network from Airbus test configuration~\cite{charara2006methodsAFDX}. Each connection represents a potential link and the boxes with $S$ combined with a number represent switches.}
    \label{fig: industrial network}
\end{figure}

Saihu is applied to an industrial-size network topology provided by \textit{Airbus} as a test configuration~\cite{tabatabaee2021deficit, charara2006methodsAFDX}; please see Figure~\ref{fig: industrial network}.

\begin{figure*}
\centering
\begin{subfigure}[b]{0.5\textwidth}
    \centering
    \includegraphics[width=\linewidth]{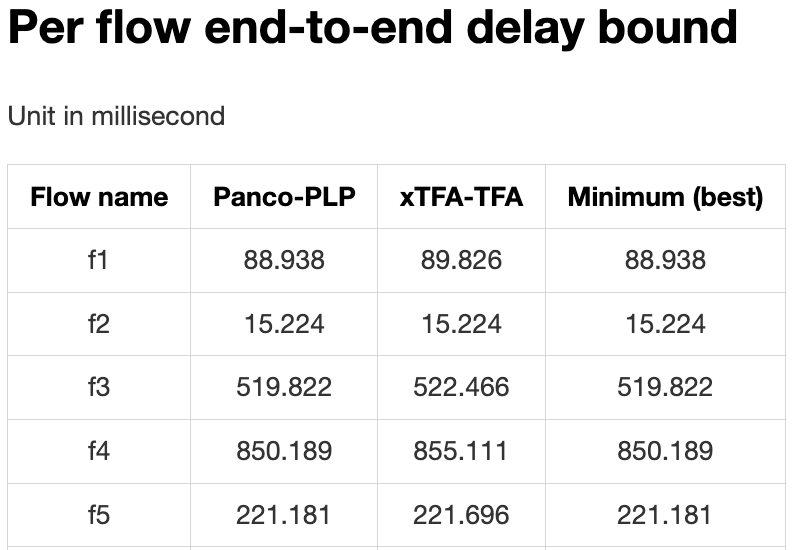}
    \caption{Flow end-to-end delay (only first 5 flows)}
    \label{fig: industrial e2e delay}
\end{subfigure}
\hfill
\begin{subfigure}[b]{0.4\textwidth}
    \centering
    \includegraphics[width=0.8\linewidth]{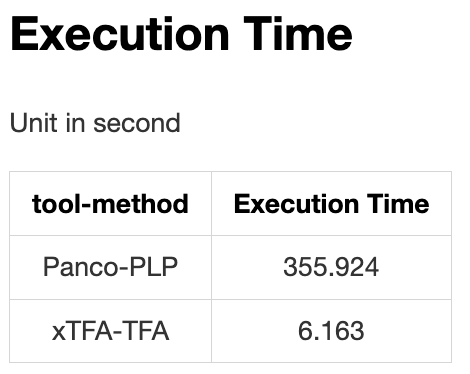}
    \caption{Execution time}
    \label{fig: industrial time}
\end{subfigure}
\caption{Example report from the topology of Figure~\ref{fig: industrial network} with 120 randomly generated flows.}
\label{fig: industrial report}
\end{figure*}

We tested the network by randomly generating $120$ flows within this network. The parameters are selected uniformly randomly from the range shown in Table~\ref{tab: industrial experiment parameters}; see Figure~\ref{fig: industrial report}. Due to the limited space here, we present the first $5$~flows out of~$120$. We only apply the methods xTFA and PLP to this network because of the cyclic dependencies.

As the number of flows increases, the complexity of PLP grows significantly. We understand real-world networks often contain thousands of flows, but we are constrained by the limitation of our computational resources. Nevertheless, Saihu provides a convenient interface for researchers. Even when the analysis options are limited by the network characteristics or the computational resources, Saihu helps researchers apply a suitable analysis method to their network of interest with relative ease.

\begin{table}[htb]
    \centering
    \footnotesize
    \begin{tabular}{c|c|c|c|c|c|c}
        & burst & arrival rate & max packet size & latency & service rate & capacity \\
        \hline
        min & 10 bytes & 64 bps & 128 bytes & 2 $\mu s$ & 1 Mbps & 256 Mbps \\
        \hline
        max & 1024 bytes & 1 kbps & 128 bytes & 200 ms & 100 Mbps & 256 Mbps
    \end{tabular}
    \caption{Parameters of the example Airbus network with size~$120$. The parameters of each flow and node is selected from the range uniformly randomly.}
    \label{tab: industrial experiment parameters}
\end{table}

